\documentclass{article}
\usepackage{amssymb,amsthm,amsmath,latexsym}
\newtheorem{thm}{{\bf Theorem}}[section]

\newtheorem{lem}{{\bf Lemma}}[section]
\newtheorem{remk}{{\bf Remark}}
\newtheorem{cor}{{\bf Corollary}}
\newtheorem{exam}{{\bf Example}}
\newtheorem{defi}{{\bf Definition}}[section]
\numberwithin{equation}{section}
\newcommand{\pf}{{\bf Proof. \ }}

\font\msbm=msbm10 at 12pt
\newcommand{\ZZ}{\mbox{\msbm Z}}
\newcommand{\RR}{{\Bbb R}}
\newcommand{\BC}{{\Bbb C}}
\newcommand{\FF}{\mbox{\msbm F}}

\begin{document}

\title
{Homogeneous Weights and M\"obius Functions on Finite Rings
   \thanks{Supported by NSFC through Grant No. 10871079.}
}
\author{Yun Fan,~~ Hongwei Liu\\
 \small  Dept of Mathematics, Central China Normal University,
   Wuhan, 430079, China
 }
\date{}

\maketitle

\insert\footins{\footnotesize
 E-mail address: yunfan02@yahoo.com.cn (Yun Fan),\quad
  h\_w\_liu@yahoo.com.cn (Hongwei Liu).}

\begin{abstract}
The homogeneous weights and the M\"obius functions and Euler
phi-functions on finite rings are discussed; some computational
formulas for these functions on finite principal ideal rings are
characterized; for the residue rings of integers, they are reduced
to the classical number-theoretical M\"obius functions and the
classical number-theoretical Euler phi-functions.

\medskip\hangindent20mm
{\it Keywords:}~ Finite ring, finite principal ideal ring,
    homogeneous weight, M\"{o}bius function, Euler phi-function.

\end{abstract}

\section{{\bf Introduction}}

The homogeneous weight on finite rings is a generalization
of the Hamming weight on finite fields and the
Lee weight on the residue ring of integers modulo $4$.
Constantinescu and Heise \cite{Con-Heise} introduced
the homogeneous weight on finite rings.
Soon after, Greferath and Schmidt \cite{G-S1}
proved the existence and uniqueness of the homogeneous weight on
a finite ring, and exhibited a formula in terms of M\"{o}bius function
and Euler phi-function to calculate the homogeneous weight,
see Eqn (\ref{homoweight-mu}) below for details.
With the generating character, Honold \cite{Honold} showed another formula
to calculate the homogeneous weight on a finite Frobenius ring.
In~\cite{V-W}, with the help of the formula in~\cite{G-S1},
 Voloch and Walker showed how to calculate the homogeneous weight
on a finite Galois ring, and estimated the homogeneous weight
of a algebraic geometry code.
There are many related references, e.g.
\cite{B-G-H}, \cite{G-O}, \cite{G-M-O}, \cite{G-S}.

However, it is not easy to obtain the values of
the M\"obius function and the Euler phi-function on a finite ring.
In this paper we exhibit precise formulas to compute
the M\"obius function and the Euler phi-function on a
finite principal ideal ring, which are similar to the
classical results on the number-theoretical M\"obius function
and Euler phi-function; hence an explicit formula of the
homogeneous weight on finite principal ideal rings is obtained.
As an application, an explicit formula of the
homogeneous weight on the residue rings of integers
is formulated in terms of the classical number-theoretical
M\"obius function and Euler phi-function.
In addition, we show a short proof of the result of \cite{Honold}
on the generating character and the homogeneous weight.

\section{Homogeneous weights, M\"obius functions and Euler phi-functions}

In this section we make some preparations.

Let $R$ be a finite ring with identity $1\ne 0$; and
$R^{\times}$ stand for the multiplication group
consisting of all units (i.e. invertible elements) of $R$.
For $x\in R$, set $Rx=\{rx\,|\, r\in R\}$ which is the left principal ideal
of $R$ generated by $x$.
Let ${\cal P}=\{Rx\mid x\in R\}$ be the set of all left principal ideals of $R$.
Obviously, $({\cal P},\supseteq)$ is a poset, where ``$\supseteq$''
is the inclusion relation.
By $|X|$ we denote the cardinality of a set $X$.
By $\RR$ we denote the set of reals.
The following is introduced in \cite{Con-Heise}.

\begin{defi}\label{defi-1}~ \rm
A map $w:R\to \RR$ is called a {\em left homogeneous weight} if $w(0)=0$
and the following two hold:

(H1)~ for $x,y\in R$, if $Rx=Ry$ then $w(x)=w(y)$;

(H2)~ there is a non-negative $\lambda\in \RR$ such that for any
non-zero $x\in R$ we have that $\sum\limits_{y\in Rx}w(y)=\lambda|Rx|$.
\end{defi}

The condition (H2) implies that the average of the homogeneous weights
on a non-zero left principal ideal is $\lambda$, which is
independent of the choice of the left principal ideal.

Similarly, one can define {\em right homogeneous weights},
which coincide with left homogeneous weights once $R$ is commutative.
In the following, for convenience,
the words ``ideal'' and ``homogeneous weight'' without further attributives
stand for left ones.

For $x,y\in R$, if $rx=y$ for some $r\in R$ then we say that
$x$ divides $y$ and denote it by $x|y$.
If $x|y$ and $y|x$ (or equivalently, $Rx=Ry$),
then we write $x\buildrel g\over\sim y$.
Clearly, ``$\buildrel g\over\sim$'' is an equivalence relation on $R$,
we call it the {\em association relation} on $R$;
the equivalence class of the association relation is called
the~{\em association class}; cf. \cite[p136]{Jacobson}.
By $\tilde x$ we denote the association class of $x$.
Because $R$ is a finite ring, one can check that
$\tilde x=R^\times x=\{ux\mid u\in R^\times\}$.
By definition, the association class $\tilde x$ is just
the set of all the principal generators of the principal ideal $Rx$.
Thus we make the following definition.
\begin{defi}\label{phiI-def}~ \rm For $x\in R$,
we set $\varphi(x)=|\tilde x|=\big|\{y\in R\mid Ry=Rx\}\big|$;
and call $\varphi(x)$ the {\em Euler phi-function} on $R$.
\end{defi}

Let ${\cal A}:=\{\tilde x\mid x\in R\}$.
If $x\,|\,y$, then for any $x'\in\tilde x$ and $y'\in\tilde y$ we have
$x'\,|\,y'$; so we can write ``$\tilde x\,|\,\tilde y$''.
Then $({\cal A},``\,|\,")$ is a poset and
\begin{equation}\label{association-pideal}
 {\cal A} ~\buildrel\cong\over\longrightarrow~ {\cal P}\,,\quad
  \tilde x~ \longmapsto~ Rx\,,
\end{equation}
is an isomorphism of posets.
Thus, Euler phi-function $\varphi$ induces a function,
denoted by $\varphi$ again, defined on ${\cal P}$:
for any $I\in{\cal P}$,
$\varphi(I)=\varphi(x)$ where $x\in I$ such that $I=Rx$;~ in other words,
$\varphi(I)$ is the number of the principal generators
of the principal ideal $I$.
Obviously, any ideal $I$ of $R$ is partitioned into disjoint union
of association classes which are contained in $I$;
by (\ref{association-pideal}), any association class corresponds to
exactly one principal ideal; so we have:
$$
 |I|=\sum_{J\in{\cal P},\,I\supseteq J}\varphi(J)\,,
 \qquad \forall~ I\in{\cal P}\,.
$$
According to the M\"obius inversion on the poset ${\cal P}$
(see~\cite[\S8.6]{Jacobson}), we have
\begin{equation}\label{phi-mu}
 \varphi(I)=\sum_{J\in{\cal P},\,I\supseteq J}\mu(I,J)\cdot|J|\,,
  \qquad \forall~ I\in{\cal P}\,,
\end{equation}
where $\mu(I,J)$ is the M\"obius function on the poset ${\cal P}$
(hence on the poset ${\cal A}$).

Greferath and Schmidt \cite{G-S1} give the following formula:
the homogeneous weight $w(x)$ of $x\in R$ is as follows
\begin{equation}\label{homoweight-mu}
w(x)=\lambda\left(1-\frac{\mu(Rx,0)}{\varphi(x)}\right)\,.
\end{equation}

\begin{remk}~ \rm
For later use,
we sketch a short proof of the formula (\ref{homoweight-mu}).
The condition (H1) of Definition \ref{defi-1} says in fact that
the homogeneous weight $w$ is a function defined on ${\cal A}\cong{\cal P}$:
for $I\in{\cal P}$ set $w(I)=w(x)$ with $x\in I$ such that $I=Rx$.
Thus we can rewrite the condition (H2) of Definition \ref{defi-1} as follows:
$$\sum\limits_{J\in{\cal P},\;I\supseteq J}\varphi(J)w(J)
 =\begin{cases}\lambda|I| & I\ne 0;\\
               0, & I=0. \end{cases}$$
Define a function $t$ on ${\cal P}$ as
$$
 t(I)=\begin{cases}\lambda|I|,& I\ne 0;\\
           0, & I=0;\end{cases}
$$
then
$$
 t(I)=\sum_{J\in{\cal P},\;I\supseteq J}\varphi(J)w(J).
  \qquad \forall~ I\in{\cal P}\,.
$$
By the M\"obius inversion on the poset ${\cal P}$, we have
$$
 \varphi(I)w(I)=\sum_{J\in{\cal P},\;I\supseteq J}\mu(I,J) t(J)\,,
   \qquad \forall~ I\in{\cal P}\,;
$$
i.e.
$$
 w(I)={1\over{ \varphi(I)}}
  \sum\limits_{J\in{\cal P},\;I\supseteq J}\mu(I,J)t(J)\,,
    \qquad \forall~ I\in{\cal P}\,.
$$
Observing the definition of the function $t(J)$,
and noting that $|0|=1$ where $0$ stands for the zero ideal,
by Eqn (\ref{phi-mu}) we have the following computation:
\begin{eqnarray*}
 w(I)&=&\frac{1}{\varphi(I)}
  \sum\limits_{0\ne J\in{\cal P},\;J\subseteq I}\mu(I,J)\cdot\lambda|J|
  =-\lambda\frac{\mu(I,0)}{\varphi(I)}
  +\frac{\lambda}{\varphi(I)}
  \sum\limits_{J\in{\cal P},\;I\supseteq J}\mu(I,J)\cdot|J| \\
 &=&-\lambda\frac{\mu(I,0)}{\varphi(I)}+\frac{\lambda}{\varphi(I)}\cdot\varphi(I)
 =\lambda\left(1-\frac{\mu(I,0)}{\varphi(I)}\right)\,;
\end{eqnarray*}
that is the formula (\ref{homoweight-mu}).
\end{remk}

\section{Generating characters of finite Frobenius rings
 and M\"obius functions}

In this section we show a link between M\"obius functions
and generating characters on finite Frobenius rings,
then deduce the Honold's formula for homogeneous weights in \cite{Honold}.

Recall that, for a finite additive group $A$,
any homomorphism $\chi:A\to\BC^\times$ is called a character of $A$,
where $\BC^\times$ is the multiplicative group of the complex field;
at that case, the restriction $\chi|_B$ of $\chi$ to any subgroup $B$ of $A$
is of course a character of $B$, called the {\em restricted character}
to $B$. The homomorphism mapping any $a\in A$ to $1\in\BC^times$
is said to be the {\em unity character} and is denoted by $1$.

For a finite ring $R$, any character of the additive group of the ring $R$
is also called a character of the ring $R$.
A character $\chi: R\to\BC^\times$ is said to be a {\em generating character}
if the restriction $\chi|_{I}$ to any non-zero principal ideal $I\in{\cal P}$
is not the unity character.

Wood~\cite{Wood} proved that a finite ring $R$ is a Frobenius ring
if and only if $R$ has a generating character.
We show that the M\"obius function is related to generating characters.

\begin{lem}~ Let $R$ be a finite Frobenius ring and
$\chi$ be a generating character of $R$. Then for any $x\in R$ we have:
\begin{equation}
 \mu(Rx,0)=\sum_{y\in\widetilde x}\chi(y)\,.
\end{equation}
\end{lem}

\pf
We prove it by induction on the cardinality $|Rx|$.
If $|Rx|=1$, then $Rx=0$ is the zero-ideal and $x=0$,
hence $\chi(0)=1=\mu(0,0)$. In the following we assume
that $x\ne 0$ and set $I=Rx$.
For $J\in{\cal P}$, we denote the set of principal generators of $J$
by $\widetilde J$, i.e. $\widetilde J=\{y\in J\mid Ry=J\}$, which
is just the corresponding association class of $J$,
see Eqn~(\ref{association-pideal}).
Note that $I$ is the disjoin union of association classes contained in $I$.
So
$$
 \sum_{y\in I}\chi(y)
 =\sum_{J\in{\cal P},\,I\supseteq J}~\sum_{y\in\widetilde J}\chi(y)
 =\sum_{y\in\widetilde I}\chi(y)+
  \sum_{J\in{\cal P},\,I\supsetneqq J}~\sum_{y\in\widetilde J}\chi(y)\,.
$$
Since the restriction $\chi|_I$ is a non-unity character of $I$,
we get $\sum_{y\in I}\chi(y)=0$.
By induction, for $J\in{\cal P}$ with $J\subsetneqq I$,
we have $\sum_{y\in\tilde J}\chi(y)=\mu(J,0)$. Thus
$$
\sum_{y\in\widetilde I}\chi(y)+
  \sum_{J\in{\cal P},\,I\supsetneqq J}~\mu(J,0)=0\,.
$$
On the other hand, as $I\ne 0$, in the partial order interval $[I,0]$
we have
$$
 \mu(I,0)+\sum_{J\in{\cal P},\,I\supsetneqq J}~\mu(J,0)
 =\sum_{J\in{\cal P},\,I\supseteq J}~\mu(J,0) =0\,.
$$
Comparing the the above two equalities, we obtaine
$$
 \sum_{y\in\widetilde I}\chi(y)=\mu(I,0)\,.\qed
$$

It follows at once from the above lemma and Eqn (\ref{homoweight-mu}) that

\begin{cor}~
Let $R$ be a finite Frobenius ring, $\chi$ be a generating character of $R$,
and $w$ be the homogenous weight on $R$ with average weight $\lambda$.
Then for any $x\in R$ we have:
\begin{equation}\label{gen-character hom-weight}
w(x)=\lambda\left(1-\frac{1}{\varphi(x)}\sum_{y\in\widetilde x}\chi(y)\right)\,.
\qed
\end{equation}
\end{cor}

We deduce Honold's formula in~\cite{Honold} from
Eqn (\ref{gen-character hom-weight}).
Recall that the associaltion class
$\tilde x=R^\times x=\{ux\mid u\in R^\times\}$, i.e.
the multiplicative group $R^\times$ acts by left translation on $\tilde x$
transitively. We denote the stable subgroup of $x$ in $R^\times$
by $R^\times_x$. Then $\tilde x\cong R^\times/R^\times_x$. So
$$|R^\times|=|R^\times :R^\times_x|\cdot|R^\times_x|
  =|\tilde x|\cdot|R^\times_x|=\varphi(x)\cdot|R^\times_x|\,;$$
and
$$
 \sum_{u\in R^\times}\chi(ux)=\sum_{v\in R^\times/R^\times_x}\,
 \sum_{u\in vR^\times_x}\chi(ux)
 =\sum_{v\in R^\times/R^\times_x}|R^\times_x|\cdot\chi(vx)
 =|R^\times_x|\cdot\sum_{y\in\widetilde x}\chi(y)\,.
$$
Substituting the above expressions into Eqn (\ref{gen-character hom-weight}),
we get
$$
 w(x)=\lambda\left(1-{1\over{|R^{\times}|}}\sum\limits_{u\in
R^{\times}}\chi(ux)\right)\,,
$$
which is just the formula for homogeneous weights in~\cite{Honold}.

\section{M\"{o}bius functions and Euler phi-functions
 on finite principal ideal rings}


A finite ring is called a {\em chain ring} if all of its ideals
form a chain with respect to inclusion relation.
Let $R$ be a finite chain ring.
By definition, one can see that: $R$ has a unique maximal
ideal $\mathfrak{m}$, and the maximal ideal can be generated by one element
$\gamma$, i.e. $\mathfrak{m}=R\gamma=\{\beta\gamma\,|\,\beta\in R\}$;
there is a positive integer $e$, called the {\em nilpotent index} of $R$,
such that $\gamma^e=0$ but $\gamma^{e-1}\ne 0$; and
$$
R=R\gamma^0\supsetneqq R\gamma^1\supsetneqq\cdots \supsetneqq
R\gamma^{e-1}\supsetneqq R\gamma^e=0
$$
are all ideals of $R$. Let
$\FF=R/{\mathfrak m}=R/R\gamma$ be the residue field of $R$,
with characteristic $p$, where $p$ is a field.
Then $|\FF|=q=p^r$, $\FF^{\times}=\FF-\{0\}$ and
$|\FF^{\times}|=p^r-1$. The following two lemmas are known, see~\cite{McD}.

\begin{lem}\label{expression}~
Let notation be as above. For any $0\ne r\in R$, there is a unique integer
$i,\, 0\le i< e$ such that $r=u\gamma^i$, where $u\in R^\times$ is a unit,
which is unique modulo $\gamma^{e-i}$. \qed
\end{lem}

\begin{lem}\label{Norton-Salagean}~ Let notation be as above.
Let $V\subseteq R$ be a set of representatives of $R$ modulo $R\gamma$.
Then

(i)~ for any $r\in R$ there are unique $r_0,\cdots,r_{e-1}\in V$ such that
$r=\sum_{i=0}^{e-1}r_i\gamma^i$;

(ii)~ $|V|=|\FF|$;

(iii)~ $|R\gamma^j|=|\FF|^{e-j}$ where $0\le j\le e-1$. \qed
\end{lem}

By Lemma~\ref{Norton-Salagean}, we have the cardinality of the chain ring
$R$ as follows.
\begin{equation}\label{cardinality-of-R}
|R|=|\FF|\cdot|R\gamma|=|\FF|\cdot |\FF|^{e-1}=|\FF|^e=q^{e}.
\end{equation}

\medskip
From now on we always assume that
\underline{$R$ is a finite principal ideal ring}.
Then we have finite chain rings $R_1,R_2,\cdots, R_s$
and an isomorphism of rings:
\begin{equation}\label{principal-ideal-ring-iso}
R\cong R_1\times R_2\times \cdots\times R_s.
\end{equation}
Let $R_k\gamma_k$ for $k=1,\cdots, s$ be the unique maximal ideal
of $R_k$ generated by $\gamma_k$ with nilpotent index $e_k$,
and $\FF_{q_k}=R_k/R_k\gamma_k$ be the residue field of $R_k$
with $q_k$ elements, where $q_k$ is a power of a prime $p_k$.
Let $({\cal P}_k,\subseteq)$ stand for the poset of all ideals of $R_k$
which is a chain as follows:
$$
 R_k=R_k\gamma_k^{0}\supsetneqq R_k\gamma_k^1 \supsetneqq
  \cdots \supsetneqq R_k\gamma_k^{e_k-1}\supsetneqq R_k\gamma_k^{e_k}=0\,.
$$

The integral interval $[0,e_k]=\{0,1,\cdots,e_k-1,e_k\}$ is also a chain:~
$0\lneqq 1\lneqq \cdots\lneqq e_{k-1}\lneqq e_k$; and the following map
\begin{equation}\label{k-anti}
 [0,e_k]\longrightarrow{\cal P}_k\,,\quad i_k\longmapsto R_k\gamma_k^{i_k}\,,
\end{equation}
is an anti-isomorphism of posets, i.e. the map is bijective and satisfies
$$ i_k\le j_k \iff R_k\gamma_k^{i_k}\supseteq R_k\gamma_k^{j_k}\,.$$
For convenience, we denote
$${\cal E}=[0,e_1]\times\cdots\times[0,e_s]\,,$$
which is the direct product of $[0,e_j]$ for $j=1,\cdots,s$, i.e.
${\bf i}\in{\cal E}$ is written as
${\bf i}=(i_1,\cdots,i_s)$ with $i_k\in[0,e_k]$, and
$$
 {\bf i}\le{\bf j}~ \iff~ i_k\le j_k,~~\forall~ k=1,\cdots,s\,.
$$

The ring isomorphism (\ref{principal-ideal-ring-iso})
induces an isomorphism of multiplicative groups:
$$
 R^\times\cong R_1^\times\times\cdots\times R_s^\times\,; \eqno(4.2^\times)
$$
for each $k$, $1\le k\le s$, take $\rho_k\in R$ such that
the image of $\rho_k$ in $R_1\times\cdots\times R_s$ is as follows
\begin{equation}\label{rho-gamma}
 \rho_k~\longmapsto~(u_1,\cdots, u_{k-1},\gamma_k,u_{k+1},\cdots,u_s)
\end{equation}
where $u_l\in R_l^\times$ provided $l\ne k$.
By the isomorphisms (\ref{principal-ideal-ring-iso}) and (4.2$^\times$)
we get the following lemma at once.

\begin{lem}\label{expression-of-elements}~ Notation as above.
Each element $x$ of $R$ can be written as
$$
 x=u\rho_1^{i_1}\cdots\rho_s^{i_s}\,, \qquad
  u\in R^\times,~~ (i_1,\cdots,i_s)
   \in{\cal E}=[0,e_1]\times\cdots\times[0,e_s]\,,
$$
where $(i_1,\cdots,i_s)$ is uniquely determined by $x$. \qed
\end{lem}

Let $x$ be as in Lemma \ref{expression-of-elements}. Then
the the image in $R_1\times\cdots\times R_s$
of the ideal $Rx$ of $R$ is as follows:
\begin{equation}\label{ideal-correspond}
 Rx~\longrightarrow~ R_1\gamma_1^{i_1}\times\cdots\times R_s\gamma_s^{i_s}\,.
\end{equation}
So each ideal $I$ of $R$ is corresponding exactly to a unique
${\bf i}=(i_1,\cdots,i_s)\in{\cal E}$ such that
$$I=R\cdot(\rho_1^{i_1}\cdots\rho_s^{i_s})\,.$$

We denote the poset of principal ideals of $R$ still by ${\cal P}$
as we did in \S2; however, at the present case ${\cal P}$ is the set of
all ideals of $R$ because $R$ is a principal ideal ring.
Following the above discussion, we see that:
$$
 {\cal E}~\longrightarrow~{\cal P},\quad {\bf i}~\longmapsto~
   R\cdot(\rho_1^{i_1}\cdots\rho_s^{i_s})\,.
$$
is a bijection; and it is an anti-isomorphism of posets
since (\ref{k-anti}) is an anti-isomorphism of posets.

For ${\bf i}=(i_1,\cdots,i_s)\in{\cal E}$, set
$\bar{\bf i}=(\bar i_1,\cdots,\bar i_s)\in{\cal E}$
where $\bar i_k=e_k-i_k$, $k=1,\cdots,s$.
Then we obtain an anti-isomorphism of posets as follows:
\begin{equation}\label{anti-E}
 {\cal E}~\longrightarrow~{\cal E},\quad
 {\bf i}=(i_1,\cdots,i_s)~\longmapsto~\bar{\bf i}=
 (\bar i_1,\cdots,\bar i_s)=(e_1-i_1,\,\cdots,\,e_s-i_s)\,.
\end{equation}
Hence we obtain the following lemma.

\begin{lem}\label{isomorphism} Let notation be as above.
The following is an isomorphism of posets:
$$
 {\cal E}~\longrightarrow~{\cal P},\quad {\bf i}~\longmapsto~
   R\cdot(\rho_1^{\bar i_1}\cdots\rho_s^{\bar i_s})\,. \eqno\Box
$$
\end{lem}\medskip

Similarly to the classical Euler phi-function in the number theory,
for $q_k$ we define a $q_k$-phi-function as follows:
$$ \varphi_k(q_k^{i_k})=\begin{cases}
  q_k^{i_k}-q_k^{i_k-1}, & i_k>0;\\
  1, & i_k=0. \end{cases}
$$
Further, we define a $q$-phi-function $\varphi$ on ${\cal E}$ by:
\begin{equation}\label{q-phi-def}
 \varphi({\bf i})=\prod_{k=1}^s\varphi_k(q_k^{i_k})
 =\prod_{1\le k\le s,\,i_k>0}(q_k^{i_k}-q_k^{i_k-1})\,.
\end{equation}

Let $x\in R$. By Definition~\ref{phiI-def}, $\varphi(x)$
is just the cardinality of the association class $\tilde x$.
By Lemma \ref{expression-of-elements} and the anti-isomorphism (\ref{anti-E}),
there is a unique ${\bf i}=(i_1,\cdots,i_s)\in{\cal E}$ such that
$x=u\rho^{\bar i_1}\cdots\rho_s^{\bar i_s}$ with $u\in R^\times$.

\begin{lem}\label{computing-of-phi}~~
$\varphi(u\rho_1^{\bar i_1}\cdots\rho_s^{\bar i_s})=\varphi({\bf
i})=
  \displaystyle\prod_{1\le k\le s,\, i_k>0}
   \left(q_k^{i_k}-q_k^{i_k-1}\right)$ .
\end{lem}

\pf By  the correspondence (\ref{ideal-correspond}),
$\varphi(u\rho_1^{i_1}\cdots\rho_s^{i_s})$ is the number of the elements
$x=(x_1,\cdots,x_s)$ such that
$R_1x_1\times\cdots\times R_sx_s
 =R_1\gamma_1^{\bar i_1}\times\cdots\times R_s\gamma_s^{\bar i_s}$
(in the product of rings $R_1\times\cdots\times R_s$),
equivalently, $\,R_kx_k=R_k\gamma_k^{\bar i_k}$ for all $k=1,\cdots,s$.
Since $R_k$ is a chain ring, $R_kx_k=R_k\gamma_k^{\bar i_k}$
if and only if $x_k\in R_k\gamma_k^{\bar i_k}-R_k\gamma_k^{\bar i_k+1}$
(set difference).
For $\bar i_k=e_k$ (equivalently, $i_k=0$),
it is obvious that $R_kx_k=R_k\gamma_k^{\bar i_k}$ if and only if $x_k=0$.
Thus
$$
\varphi(u\rho_1^{\bar i_1}\cdots\rho_s^{\bar i_s})
 = \prod_{1\le k\le s}\;
      \left|R_k\gamma_k^{\bar i_k}-R_k\gamma_k^{\bar i_k+1}\right|
 = \prod_{1\le k\le s,\,i_k>0} \left(q_k^{i_k}-q_k^{i_k-1}\right)
 =\varphi({\bf i})\,.
 \eqno\Box
$$

Recall that we denote the M\"obius function on the poset ${\cal P}$
by $\mu(I,J)$ where $I,J\in{\cal P}$.
We have the isomorphism of posets
${\cal P}\cong{\cal E}=[0,e_1]\times\cdots\times[0,e_s]$.
For ${\bf i}=(i_1,\cdots,i_s)\in{\cal E}$ and
${\bf j}=(j_1,\cdots,j_s)\in{\cal E}$, we have
the M\"obius function $\mu({\bf i},{\bf j})$.
Since the M\"obius function has product property
(see~\cite[Theorem 8.10]{Jacobson}),
we have $\mu({\bf i},{\bf j})=\prod_{k=1}^s\mu_k(i_k,j_k)$,
where $\mu_k(i,j)$ is the M\"obius function on the poset $[0,e_k]$.
Further, since $[0,e_k]$ is a chain, we have (see~\cite[Theorem 8.9]{Jacobson}):
$$
 \mu_k(i_k,j_k)=\begin{cases}1, & i_k-j_k=0;\\
                      -1, & i_k-j_k=1;\\
                       0, & \mbox{otherwise.} \end{cases}
$$
If there is an index $k$ such that $i_k-j_k<0$ or $i_k-j_k>1$,
then $\mu_k(i_k,j_k)=0$ hence $\mu({\bf i},{\bf j})=0$.
Otherwise, $0\le i_k-j_k\le 1$ for all $k=1,\cdots,s$;
and $\mu_k(i_k,j_k)$ contributes $1,-1$ to $\mu({\bf i},{\bf j})$
when $i_k-j_k=0,1$ respectively. So
\begin{equation}\label{mu-PIR}
 \mu({\bf i},{\bf j})=\begin{cases}
  0, & \mbox{if $i_k-j_k<0$ or $i_k-j_k>1$ for some index;}\\
(-1)^{\beta({\bf i},{\bf j})}, &\mbox{otherwise;}\end{cases}
\end{equation}
where $\beta({\bf i},{\bf j})=\left|\{k\mid i_k-j_k=1\}\right|$,
the number of indexes $k$ such that $i_k-j_k=1$.

\begin{lem}\label{computing-of-mu}~~
Let $I,J\in{\cal P}$ correspond ${\bf i},{\bf j}\in {\cal E}$, respectively,
under the isomorphism in Lemma \ref{isomorphism}, i.e.
$I=R\cdot(\rho_1^{\bar i_1}\cdots\rho_s^{\bar i_s})$,
$J=R\cdot(\rho_1^{\bar j_1}\cdots\rho_s^{\bar j_s})$. Then
$$
 \mu(I,J)=\mu({\bf i},{\bf j})=\begin{cases}
  0, &\mbox{if $i_k-j_k<0$ or $i_k-j_k>1$ for some index;}\cr
(-1)^{\beta({\bf i},{\bf j})}, & \mbox{otherwise.}\end{cases}
$$
where $\beta({\bf i},{\bf j})$
is the number of indexes $k$ such that $i_k-j_k=1$.
\end{lem}

\pf
It follows from Lemma~\ref{isomorphism} and Eqn~(\ref{mu-PIR}) immediately.
\qed

\medskip
Noting that, by Lemma~\ref{expression-of-elements},
any element of $R$ is written as
$u\rho_1^{\bar i_1}\cdots\rho_s^{\bar i_s}$ with $u\in R^\times$
and ${\bf i}=(i_1,\cdots,i_s)\in[0,e_1]\times\cdots\times[0,e_s]$,
we get a precise formula to compute the homogeneous weight $w(-)$
on the finite principal ideal ring $R$ as follows.

\begin{thm}\label{Homo-PIR}~ Let notation be as above. Then
$$
w\big(u\rho_1^{\bar i_1}\cdots\rho_s^{\bar i_s}\big)
 =\lambda\left(1-\frac{\mu({\bf i},{\bf 0})}{\varphi({\bf i})}\right)\,,
$$
where $\mu({\bf i},{\bf 0})$ is computed in Lemma~\ref{computing-of-mu}
and $\varphi({\bf i})$ is computed in Lemma~\ref{computing-of-phi}.\qed
\end{thm}

\medskip

As a special case where $s=1$, we have the following corollary at once.

\begin{cor}~
If $R$ is a finite chain ring with a unique maximal ideal $R\gamma$ and
nilpotent index $e$, then for any $x\in R$ we have
$$
w(x)=\left\{\begin{array}{ll} 0, &  x=0;\\
{\lambda q\over {q-1}},~ &  0\ne x\in R\gamma^{e-1};\\
\lambda, \, &\mbox{otherwise};\\ \end{array}\right.
\qquad\mbox{where}~~q=|R/R\gamma|. \eqno\Box
$$
\end{cor}

\pf Write $x=u\gamma^{\bar i}$ with $u\in R^\times$ and $i\in[0,e]$,
where $\bar i=e-i$. By Theorem~\ref{Homo-PIR} we have
$$
w(x)=w(u\gamma^{\bar i})=\lambda\left(1-\frac{\mu(i,0)}{\varphi(i)}\right)\,.
$$
If $x=0$, i.e. $\bar i=e$ hence $i=0$,
then $\varphi(0)=1=\mu(0,0)$; so $w(x)=0$.

if $0\ne x\in R\gamma^{e-1}$, i.e. $i=1$,
then, by Lemma~\ref{computing-of-phi}, $\varphi(1)=q-1$;
by Lemma~\ref{computing-of-mu}, $\mu(i,0)=-1$; so
$$w(x)=\lambda\left(1-\frac{-1}{q-1}\right)=\frac{\lambda q}{q-1}\,.$$

Otherwise, $e\ge i\ge 2$, then $\mu(i,0)=0$, hence $w(x)=\lambda(1-0)=\lambda$.
\qed

\section{The homogeneous weight on the residue rings of integers}

In this section we reformulate the computation of
the homogeneous weight on the residue rings of integers
with the classical M\"obius function and the Euler phi-function
in the number theory.

For any integer $m>1$ we have a standard decomposition
$m= p_1^{i_1}\cdots p_r^{i_r}$, where $p_1$, $\cdots$, $p_r$ are
distinct primes and $i_k>0$ for $k=1,\cdots,r$.
The classical number-theoretical Euler phi-function is:
$$
 \varphi(m)=\prod_{1\le k\le r}\left(p_k^{i_k}-p_k^{i_k-1}\right)\,;
$$
and the classical number-theoretical M\"obius function is:
$$
 \mu(m)=\begin{cases}(-1)^r, & \mbox{if $i_k=1$ for all $k=1,\cdots,r$;}\\
  0, &\mbox{otherwise.}\end{cases}
$$
And $\varphi(1)=\mu(1)=1$.

\begin{thm}\label{Hom-Residue-ring}~
Let $n$ be a positive integer, $w$ be the homogeneous weight on ${\Bbb Z}_n$.
Then any element of ${\Bbb Z}_n$ can written as
$u\cdot n/m$ where $u\in{\Bbb Z}_n^\times$ and $m|n$, and
$$
 w(u\cdot n/m)=w(n/m)=\lambda\left(1-\frac{\mu(m)}{\varphi(m)}\right)\,.
$$
\end{thm}

\pf Let $n= p_1^{e_1}\cdots p_s^{e_s}$ where $p_1$, $\cdots$, $p_s$
are distinct primes and $e_k>0$ for all $k=1,\cdots,s$. Then
$$
 {\Bbb Z}_n\cong {\Bbb Z}_{p_1^{e_1}}\times\cdots\times {\Bbb Z}_{p_s^{e_s}}\,,
$$
where ${\Bbb Z}_{p_k^{e_k}}$ is a chain ring with unique maximal ideal
$p_k{\Bbb Z}_{p_k^{e_k}}$, nilpotent index $e_k$ and residue filed
isomorphic to $\ZZ_{p_k}$ of order $p_k$.
The element $p_k\in{\Bbb  Z}_{n}$ is mapped by the above isomorphism to:
$$
 p_k~\longmapsto~(p_k,\cdots,p_k,\cdots,p_k)\,;
$$
where the $l$'th coefficient $p_k$ in the right hand size is a unit
of ${\Bbb Z}_{p_l^{e_l}}$ provided $l\ne k$, compare with Eqn~(\ref{rho-gamma}).
Thus, by Lemma~\ref{expression-of-elements},
any element $a$ of ${\Bbb Z}_n$ is written as
$$
x=up_1^{l_1}\cdots p_s^{l_s},\qquad u\in{\Bbb Z}_n^\times,\quad
 {\bf l}=(l_1,\cdots,l_s)\in{\cal E}=[0,e_1]\times\cdots\times[0,e_s].
$$
Set $(i_1,\cdots,i_s)={\bf i}=(e_1-l_1,\,\cdots,\,e_s-l_s)$
and $m=p_1^{i_1}\cdots p_s^{i_s}$; then $x=u\cdot n/m$.
So $w(x)=w(u\cdot n/m)=w(n/m)$, by Theorem~\ref{Homo-PIR}, we have
$$
 w(n/m)=w(p_1^{\bar i_1}\cdots p_s^{\bar i_s})=
 \lambda\left(1-\frac{\mu({\bf i},{\bf 0})}{\varphi({\bf i})}\right)\,.
$$
Further, by Eqn~(\ref{q-phi-def}) we have
$$
 \varphi({\bf i})=\prod_{1\le k\le s,\,i_k>0}
   (p_k^{i_k}-p_k^{i_k-1})=\varphi(m)\,;
$$
and by Lemma~\ref{computing-of-mu} it is easy to check that
$\mu({\bf i},{\bf 0})=\mu(m)$. So
$$
 w(n/m)=\lambda\left(1-\frac{\mu(m)}{\varphi(m)}\right). \eqno\Box
$$

\begin{remk}~ \rm More precisely, for $m=p_1^{i_1}\cdots p_s^{i_s}$
where $(i_1,\cdots,i_s)\in[0,e_1]\times\cdots\times[0,e_s]$,
we have
$$
 w(n/m)=\begin{cases}\lambda,&\mbox{if there is an $i_k>1$;}\\
 \lambda(\varphi(m)-1)/\varphi(m), &
 \mbox{if every $i_k\le 1$ and $\beta(m)$ is even;}\\
 \lambda(\varphi(m)+1)/\varphi(m), &
   \mbox{if every $i_k\le 1$ and $\beta(m)$ is odd;}\end{cases}
$$
where $\beta(m)$ stands for the number $k$ such that $i_k=1$.
In particular, for $m=n$ we get that
$$
 w(1)=\begin{cases}\lambda,&\mbox{if there is an $e_k>1$;}\\
 \lambda(\varphi(n)-(-1)^s)/\varphi(n), &\mbox{otherwise.}\end{cases}
$$
\end{remk}
\begin{exam}~\rm
Applying Theorem~\ref{Hom-Residue-ring},
we calculate the Table \ref{ZZ24} of the homogeneous weights on
$\ZZ_{24}$, $\ZZ_{12}$ and $\ZZ_6$.
\end{exam}

\begin{table}[h]
 \caption{Homogeneous weights on $\ZZ_{24}, \ZZ_{12}$ and $\ZZ_6$}
\label{ZZ24}
 \begin{center}
{\scriptsize
 \begin{tabular}{|r|r|r|r|}
\noalign{\hrule height0.8pt} \multicolumn{1}{|c}{$\ZZ_{24}$} &
\multicolumn{1}{|c|}{$w(x)$} &\multicolumn{1}{|c}{$\ZZ_{24}$}
& \multicolumn{1}{|c|}{$w(x)$}\\
\hline
$0$   & $ 0$ &$12$ &$2\lambda$  \\
$1$   & $ \lambda   $   &$13$ &$\lambda$\\
$2$   & $ \lambda   $ &$14$ &$\lambda$  \\
$3$   & $ \lambda   $    &$15$ &$\lambda$\\
$4$   & $ {1\over 2}\lambda   $ &$16$ &${3\over 2}\lambda$  \\
$5$   & $ \lambda   $   &$17$ &$\lambda$ \\
$6$   & $ \lambda   $    &$18$ &$\lambda$\\
$7$   & $ \lambda   $   &$19$ &$\lambda$ \\
$8$   & $ {3\over 2}\lambda   $  &$20$ &${1\over 2}\lambda$   \\
$9$   & $ \lambda   $   &$21$ &$\lambda$  \\
$10$   & $ \lambda   $   &$22$ &$\lambda$  \\
$11$   & $ \lambda   $    &$23$ &$\lambda$ \\
\noalign{\hrule height0.8pt}
   \end{tabular}} \qquad\qquad{\scriptsize
 \begin{tabular}{|r|r|}
\noalign{\hrule height0.8pt} \multicolumn{1}{|c}{$\ZZ_{12}$}
& \multicolumn{1}{|c|}{$w(x)$}\\
\hline
$0$   & $ 0   $  \\
$1$   & $ \lambda   $   \\
$2$   & $ {1\over 2}\lambda   $   \\
$3$   & $ \lambda   $    \\
$4$   & $ {3\over 2}\lambda   $   \\
$5$   & $ \lambda   $    \\
$6$   & $ 2\lambda   $    \\
$7$   & $ \lambda   $    \\
$8$   & $ {3\over 2}\lambda   $     \\
$9$   & $ \lambda   $     \\
$10$   & $ {1\over 2}\lambda   $     \\
$11$   & $ \lambda   $     \\
\noalign{\hrule height0.8pt}
   \end{tabular}} \qquad\qquad
   {\scriptsize
\begin{tabular}{|r|r|}\noalign{\hrule height0.8pt}
\multicolumn{1}{|c}{$\ZZ_6$}
& \multicolumn{1}{|c|}{$w(x)$}\\
\hline
$0$   & $ 0   $  \\[3pt]
$1$   & ${1\over 2} \lambda   $   \\[3pt]
$2$   & $ {3\over 2}\lambda   $   \\[3pt]
$3$   & $ 2\lambda   $    \\[3pt]
$4$   & $ {3\over 2}\lambda   $   \\[3pt]
$5$   & $ {1\over 2}\lambda   $    \\
\noalign{\hrule height0.8pt}
   \end{tabular}}
\end{center}
\end{table}

\section*{Acknowledgements}
The authors thank Prof Steven T. Dougherty for may helpful discussion.


\end{document}